\documentclass[
    ,final            % use final for the camera ready runs
  ]
  {aipproc}

\layoutstyle{8x11double}

\begin{document}

\title{Anomaly Mediation 
and Radius Stabilization by a Boundary Constant Superpotential in a Warped Space}

\classification{11.30.Pb, 11.25.-w, 12.60.Jv}
\keywords      {Radius stabilization, Boundary constant superpotential, Anomaly Mediation}

\author{Nobuhito Maru}{
  address={Department of Physics, Kobe University, Kobe 657-8501, Japan}
}

\author{Norisuke Sakai}{
  address={Department of Mathematics, Tokyo Woman's Christian University, Tokyo 167-8585, Japan}
}

\author{Nobuhiro Uekusa}{
  address={Department of Physics, Osaka University, Toyonaka, Osaka 560-0043, Japan}
  ,%altaddress={<author1 address>} % additional visiting address
}

\begin{abstract}
We present a very simple model of the radius stabilization in a supersymmetric (SUSY) Randall-Sundrum model 
with a hypermultiplet and a boundary constant superpotential. 
A wide range of parameters where the anomaly mediation of SUSY breaking is dominated 
is found although there are many problematic bulk effects of SUSY breaking. 
A negative cosmological constant in the radius stabilized vacuum can be cancelled 
by a localized SUSY breaking.  
Making use of this localized SUSY breaking also solves the $\mu$-problem by Giudice-Masiero mechanism. 
\end{abstract}

\maketitle

%%%%%%%%%%%%%%%%%%%%%%%%%%%%%%%%%%%%%%%%%%%%
%% MAINMATTER
%%%%%%%%%%%%%%%%%%%%%%%%%%%%%%%%%%%%%%%%%%%%

\section{Introduction}

It is well known that the main motivation of introducing extra dimensions is to solve the gauge hierarchy problem. 
However, there is an alternative motivation in SUSY brekaing model building, namely the solution to SUSY flavor problem. 
In the gravity mediation of four dimensions, sfermion masses are generated by contact interactions 
between the hidden sector fields and the minimal SUSY standard model (MSSM) fields. 
The interactions are in general flavor {\em dependent} since there is no physical symmetry reason 
to be flavor diagonal, which give rise to an excessive $K^0-\bar{K}^0$ mixing, for example. 
To suppress these interactions in four dimensions, we must introduce additional symmetries. 

On the other hand, in higher dimensional case, if the hidden sector and MSSM sector are separated 
along the extra dimensions, the above mentioned contact interactions for sfermion masses 
are forbidden by locality. 
Then, the dominant sfermion mass is generated by anomaly mediation without SUSY flavor problem. 

However, this is not the end of the story. 
The radius of the compactified dimensions must be stabilized. 
Although the nontrivial radion potential is generated once SUSY is broken, 
it seems not to be stabilized by only the gravity multiplet in the bulk. 
Thus, we must introduce additional bulk fields to stabilize the radius. 
In that situation, we have to check whether these additional bulk fields 
does not generate flavor violating sfermion masses. 

In this talk, we present a very simple model of the radius stabilization and 
anomaly mediation dominated SUSY breaking. 

\section{Model}

%\subsection{<A subsection>}
We consider a five-dimensional SUSY model on the Randall-Sundrum
background, whose metric is
\begin{eqnarray}
ds^2 = e^{-2R \sigma} \eta_{\mu\nu} dx^\mu dx^\nu + R^2 dy^2,
\quad
\sigma(y)\equiv k|y|,
 \label{metric}
\end{eqnarray}
where $\eta_{\mu\nu}={\rm diag}(-1,+1,+1,+1)$,
$R$ is the radius of $S^1$ of the orbifold $S^1/Z_2$,
$k$ is the $AdS_5$ curvature scale, and the angle of $S^1$
is denoted by $y~(0 \le y \le \pi)$. 

As a minimal model to break SUSY %the supersymetry 
and to stabilize the radius, we introduce a single hypermultiplet. 
In terms of superfields for four manifest SUSY, %supersymmetry, 
the single hypermultiplet is represented 
by chiral supermultiplets $\Phi, \Phi^c$, and 
our Lagrangian reads 
\begin{eqnarray}
{\cal L}_5 &=& \int d^4 \theta
\frac{1}{2} \varphi^\dag \varphi (T+T^\dag)
e^{-(T+T^\dag)\sigma} \nonumber \\
&& \times (\Phi^\dag \Phi + \Phi^c \Phi^{c\dag} - 6M_5^3)
\nonumber \\
&+& \int d^2 \theta
\left[
\varphi^3 e^{-3T \sigma} \bigg\{
\Phi^c \left[
\partial_y - \left( \frac{3}{2} - c \right)T \sigma'
\right] \Phi \right. \nonumber \\
&& \left. + W_b
\bigg\} + {\rm h.c.}
\right] ,
\label{lagrangian}
\end{eqnarray}
where the compensator chiral supermultiplet $\varphi$ 
(of supergravity), and the radion chiral supermultiplet 
$T$ are denoted as 
$\varphi = 1 + \theta^2 F_{\varphi}$ and
$T=R + \theta^2 F_T$,
respectively. 
 
%and the chiral supermultiplets representing
%the hypermultiplet is denoted as
%$\Phi, \Phi^c$.
The $Z_2$ parity is assigned to be even (odd) for
$\Phi (\Phi^c)$.
The derivative with respect to $y$ is denoted by $'$,
such as $\sigma'\equiv d\sigma/dy$.
The five-(four-) dimensional Planck mass is denoted as $M_5$ ($M_4$).
%$c$ is a 
A bulk mass parameter for the hypermultiplet is denoted as $c$.  
 
Here we consider a model with a constant (field independent)
superpotential localized at the fixed point $y=0$
\begin{eqnarray}
W_b \equiv 2M_5^3 w_0 \delta(y) ,
\label{eq:boundary_pot}
\end{eqnarray}
where $w_{0}$ is a dimensionless constant.

\subsection{Radius stabilization}
The background solutions of equations of motion for the hyperscalars
at the leading order of $w_0$ are given by
\begin{eqnarray}
%&&
\phi(y)&=&N_2\exp\left[\left({3\over 2}-c\right)R\sigma\right],
 \label{phin2}
\\
\phi^c(y)&=&\hat{\epsilon}(y)
   \left({\phi^\dagger\phi\over 6M_5^3}-1\right)^{-1}
 \left({\phi^\dagger\phi\over 6M_5^3}\right)^{{5/2-c\over 3-2c}} \times \nonumber \\
&& \left[c_1 +c_2
 \left({\phi^\dagger\phi\over 6M_5^3}\right)^{-{1-2c\over 3-2c}}
\left({\phi^\dagger\phi\over 6M_5^3}+{2\over 1-2c}\right)\right]
\nonumber
\\
\label{phichic}
\end{eqnarray}
where $c\neq 1/2, 3/2$, and $\hat{\epsilon(y)}$ is a sign function of $y$. 
%\begin{eqnarray}
%\hat{\epsilon}(y) \equiv
%\left\{
%\begin{array}{cc}
%+1, & 0<y<\pi \\
%-1, & -\pi<y<0
%\end{array}
%\right.
%.
%\label{eq:sign_function}
%\end{eqnarray}
The solution has three complex integration constants:
$c_1, c_2$ are the coefficients of two independent
solutions for $\phi^c$, and the overall complex constant
$N_2$ for the flat direction $\phi$.
Two of these three complex integration constants
are determined by the boundary conditions.
The single remaining constant (which we choose as $N_2$)
is determined by the potential minimization. 
%(stabilization).

\begin{figure}
  \includegraphics[height=.2\textheight]{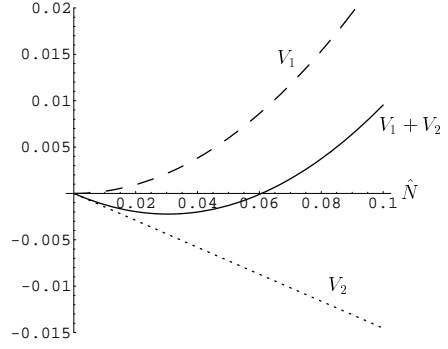}
  \caption{Radion potential}
  \label{potential}
\end{figure}

With the backgrounds (\ref{phin2}) and (\ref{phichic}),
the potential is obtained as \cite{MSU1}
\begin{eqnarray}
 V
&=& {3M_5^3 k w_0^2\over 2} \times \nonumber \\
&& \left\{
  \frac{-2(1-2c)\hat{N}^{4-2c-\frac{1}{3-2c}}}{(1-2c)(e^{2Rk\pi}-1)\hat{N}
+2(e^{(2c-1)Rk\pi}-1)}
%\hat{N}^{4-2c-\frac{1}{3-2c}}
\right. \nonumber \\
&& \left. +\frac{\hat{N}}{1-\hat{N}}
\left( -4c^2+12c-6 +\frac{3-2c}{3(1-\hat{N})}
\right)
  \right\}.
\label{potentialwp0}
\end{eqnarray}
where 
a %the 
dimensionless quantity is defined as $\hat{N}\equiv |N_2|^2/(6M_5^3)$.
We need to require the stationary condition for both modes $R$ and
$N_2$, namely $\partial V/\partial R=0$ and $\partial V/\partial \hat{N}=0$.
From these stationary conditions,
we find that there is a unique nontrivial minimum with 
finite values of the radius $R$ and of 
the normalization $N_2$ for the
flat direction $\phi$ provided 
$c < c_{\rm cr} \equiv {17-\sqrt{109}\over 12}.$
%At the critical value of the mass parameter $c_{\rm cr}$,
%the minimum occurs at the infinite radius and vanishingnormalization $N_2$,
%$\hat{N}(c_{{\rm cr}})=0$, $R(c_{{\rm cr}})=\infty$.
To examine the stabilization 
%for $c < c_{\rm cr}$ 
in more detail,
%closely,
we parameterize $c=c_{{\rm cr}}-\Delta c$ with a small $\Delta c$. 
After using the stationary condition solution 
$\hat{N}=e^{-(3-2c)Rk\pi}$, 
we find that the potential (\ref{potentialwp0}) shown in Fig. \ref{potential}
%as a function of $c=\ccr-\Delta c$ and $\hat{N}$ 
consists of two pieces 
at the leading order of $\Delta c \equiv c_{{\rm cr}}-c$ and $\hat{N}$ 
\begin{eqnarray}
  V  &\approx& {3M_5^3 kw_0^2\over 2}(V_1 +V_2) ,
\\
  V_1 &\equiv& \frac{ 2 (2 c_{{\rm cr}}-1) }{ 3-2 c_{{\rm cr}} }
    \hat{N}^{ \frac{4 c_{{\rm cr}}^2-12 c_{{\rm cr}} + 10 }{3-2 c_{{\rm cr}} } },
\\
  V_2&\equiv& -\hat{N} \left(-8 c_{{\rm cr}} + {34\over 3} \right) \Delta c.
\end{eqnarray}
%The potential $V$ and its pieces $V_1, V_2$ are depicted
%as a function of $\hat N$
%in Fig.\ref{fig:1}.
%\begin{figure}[htbp]
%\begin{center}
%\includegraphics[width=0.3\textwidth,angle=0]{stabiliz.eps}
%\caption{Potential for $c=c_{\textrm{\scriptsize cr}}-\Delta c$}
%\label{fig:1}
%\end{center}
%\end{figure}
%It is now obvious that a unique minimum occurs at
%finite values of $\hat N$ provided $\Delta c >0$
%($c < c_{\rm cr}$) and that the minimum point approaches
%$\hat N \to 0$ as $\Delta c \to 0$
%($c \to c_{\rm cr}$).  
%Actually the Fig.\ref{fig:1} demonstrates
%If we eliminate $R$ by using the stationary condition,
%we need to obtain a minimum along the direction of $\hat N$.
%There is only the stability along the direction of $\hat N$,
%after the other variable $R$ is eliminated by the
%stationary condition. %(\ref{dvr0}).
%We have checked that this minimum point gives
%a true minimum of the potential $V(R, \hat N)$
%as a function of two variables 
%$R$ and $\hat N$, 
%establishing the stability in both directions.
The stationary point at
the leading order of $\Delta c$ is obtained as
\begin{eqnarray}
  R \approx {1\over 10k}\left(\ln {1\over \Delta c}-3.4\right) ,
\label{Rdeltac}
\end{eqnarray}
which means that the radius is stabilized with the size
of $R>1/k$ for $\Delta c<10^{-6}$.

The masses of radion and moduli field $N_2$ are found by 
diagonalizing the kinetic term and mass squared matrix, 
\begin{eqnarray}
m^2_{\rm radion}  &\approx& k^2 w_0^2 0.38 (3.4+ \ln \Delta c)^2 (\Delta c)^{1.7}, \\
m^2_{\rm moduli} &\approx& k^2 w_0^2 0.47 (\Delta c)^{0.70}
\end{eqnarray}
which are estimated to be 
\begin{eqnarray}
  m_{\rm radion} \sim  1~{\rm TeV},
\qquad
  m_{\rm moduli} \sim 100~{\rm TeV}
\end{eqnarray}
for $k w_0 \sim 10^{7}~{\rm GeV}$ and $\Delta c\sim 10^{-6}$.

At the stationary point the potential becomes
\begin{eqnarray}
  V \approx  -10^{37}(kw_0)^2(\Delta c)^{1.2} \sim -(10^{10}~{\rm GeV})^4.
  \label{eq:pst}
\end{eqnarray}
If we add a spurion supermultiplet $X=F_X\theta^2$ localized at $y=0$ 
with the Lagrangian
\begin{eqnarray}
  {\cal L}_X=\left[
     \int d^4\theta |\varphi|^2 X^\dag X
       + \left( \int d^2\theta \varphi^3 m^2 X+ \textrm{h.c.} \right) \right]
        \delta(y) , 
        \label{spuri}
\end{eqnarray}
%We can show that 
the cosmological constant
can be cancelled by an $F$ term
contribution
\begin{eqnarray}
   \sqrt{F_X} \approx 10^{10}~{\rm GeV} .
\end{eqnarray}
%which means $\sqrt{F_X}\sim 10^{10}~{\rm GeV}$ for $\Delta c \sim 10^{-6}$.

We comment that this localized $F$-term SUSY breaking can be utilized 
for solving the $\mu$-problem by Giudice-Masiero mechanism \cite{MSU3}. 
If two Higgs superfields $H_u, H_d$ in the MSSM are assumed to be localized at $y=0$, 
the following K\"ahler terms are allowed 
\begin{eqnarray}
K = \int d^4\theta |\varphi|^2 \left[ 
{X^\dag\over  M_4} H_u H_d 
  + {X^\dag X \over M_4^2} 
     H_u H_d + \textrm{h.c.} 
\right] \delta(y). 
\label{GMKahler}
\end{eqnarray}
As in the case of cancellation of the cosmological constant, 
the VEV of the scalar component for the chiral multiplet $X$ is assumed to be zero. 
This ensures that equations of motion for auxiliary fields are unchanged. 
Namely, our successful stabilization mechanism is not affected by addition of (\ref{GMKahler}). 

After SUSY breaking, the correct order of $\mu$-term and $B\mu$-term are generated 
from the first and the second terms, 
respectively.\footnote{The coefficients of each term in (\ref{GMKahler}) 
are assumed to be an order unity.}
\begin{eqnarray}
\mu^2 \sim B\mu \sim \left(\frac{F_X}{M_4} \right)^2 \sim (100~{\rm GeV})^2
\end{eqnarray}
where $\sqrt{F_X} \sim 10^{10}$ GeV, 
which is required for canceling the cosmological constant, is used. 
It is very interesting that canceling the cosmological constant and the solution to the $\mu$-problem 
are realized simultaneously by the same origin of SUSY breaking effect.

%The cosmological constant can be also cancelled by 
% a $D$ term contribution
% for SUSY breaking
%and that the contributions of these sectors to
%the soft mass and gravitino mass are small \cite{MSU2}.

\subsection{SUSY breaking mass spectrum}

%\subsubsection{Soft mass by anomaly mediation}
%In a SUSY Randall-Sundrum model,
%anomaly mediated scalar mass is given by
%$\tilde{m}_{\rm AMSB}
% \sim (g^2/ 16\pi^2)$ $\cdot$ $\langle F_\omega/\omega\rangle$
%\cite{Luty:2002ff}.
%Here the superfield $\omega$ is defined as
%a rescaled compensator multiplet $\omega=\varphi e^{-T\sigma}$
%and we denote its lowest component also as $\omega$,
%and $g$ is gauge coupling constant for visible sector fields.
We assume that the MSSM fields are localized at $y=\pi$. 
In our model, the anomaly mediated scalar mass becomes
\begin{eqnarray}
  \tilde{m}_{\rm AMSB}
 &\sim& {g^2\over 16\pi^2}  (F_\varphi-F_T\sigma)\bigg|_{y=\pi} \nonumber \\
 &\sim&  {\cal O}(10^{-4})\times g^2 k w_0 \sim 100~{\rm GeV}. 
\end{eqnarray}
where $g$ is gauge coupling constant for visible sector fields 
and $g^2 kw_0 \sim 10^6$~GeV is used to obtain the last expression. 
%Using the hyperscalar background (\ref{phin2}), (\ref{phichic})
%and the stationary condition,
% we obtain the anomaly mediated scalar mass as
%\begin{eqnarray}
% \tilde{m}_{\rm AMSB}
%  \sim  {\cal O}(10^{-4})\times g^2 k w_0 \sim 100~{\rm GeV}. 
%\end{eqnarray}
%for $g^2 kw_0\sim 10^6$~GeV. 
We can show that soft masses mediated by Kaluza-Klein modes
in our model are smaller than those by the anomaly mediation \cite{MSU2}. 
The brane-to-brane mediation of $F_X$ by a bulk gravity (a hypermultiplet), 
which is tachyonic (flavor dependent), is suppressed enough 
for $\sqrt{F_X} < 10^{11} {\rm GeV}$
comparing to the anomaly mediation \cite{MSU2}. 
Therefore our model passes the flavor-changing neutral current constraints. 

For gaugino mass, the anomaly mediation is also dominant 
as long as additional interactions with SUSY breaking gauge singlets
are not included in the visible sector gauge kinetic terms.
%The gaugino mass is of the same order as the scalar mass.

%\subsubsection{Radion and moduli masses}

%Substituting Eq.(\ref{phin2})
%into the Lagrangian %${\cal L}_{\textrm{\scriptsize kin}}$
%and diagonalizing the kinetic term and mass-squared matrix,
%we find that at the leading order of $e^{-Rk\pi}$
%the lighter physical mode is
%almost
%exclusively made of the radion
Finally, the gravitino mass is obtained by solving the equation of motion 
in the presence of the constant superpotential $w_0$, 
\begin{equation}
m_{3/2} \sim 6 w_0 k \sim 10^7~{\rm GeV}, 
\end{equation}
which is a relatively large gravitino mass specific to the SUSY Randall-Sundrum model \cite{MSU1}. 

%\subsubsection{short summary}

%Some url test \url{http://www.world.universe}.

%\subsubsection{<A subsubsection>}

%%%%%%%%%%%%%%%%%%%%%%%%%%%%%%%%%%%%%%%%%%%%
%% Sample figure:
%%
%% The option [height=...] scales the picture to the given height,
%% without it it would be printed at its nominal size
%%%%%%%%%%%%%%%%%%%%%%%%%%%%%%%%%%%%%%%%%%%%

%%%%%%%%%%%%%%%%%%%%%%%%%%%%%%%%%%%%%%%%%%%%
%% SAMPLE TABLE
%%
%% Shows the use of \tablehead and \tablenote
%% macros
%%%%%%%%%%%%%%%%%%%%%%%%%%%%%%%%%%%%%%%%%%%%

\section{Conclusions}

We have presented a very simple model of the radius stabilization and 
anomaly mediation in SUSY Randall-Sundrum model 
with a massive hypermultiplet and a boundary constant superpotential. 
We found a range of parameters where other dangerous bulk SUSY breaking mediation effects 
are suppressed to avoid the SUSY flavor problem. 
It is interesting that cancellation of the cosmological constant and 
solving the $\mu$-problem by Giudice-Masiero mechanism are simultaneously realized 
by the same localized $F$-term SUSY breaking. 

A negative slepton problem is still remained unsolved. 
We are now trying to solve this problem without spoiling our stabilization mechanism \cite{MSU3}.

%%%%%%%%%%%%%%%%%%%%%%%%%%%%%%%%%%%%%%%%%%%%%%%%
%% BACKMATTER
%%%%%%%%%%%%%%%%%%%%%%%%%%%%%%%%%%%%%%%%%%%%%%%%

\begin{theacknowledgments}
We would like to thank the organizers for providing me with an oppotunity 
to talk at the conference. 
This work is supported in part by Grant-in-Aid for
Scientific Research from the Ministry of Education,
Culture, Sports, Science and Technology, Japan No.17540237 (N.S.)
and No.18204024 (N.M. and N.S.).
\end{theacknowledgments}




%\bibliography{sample}

\IfFileExists{\jobname.bbl}{}
 {\typeout{}
  \typeout{******************************************}
  \typeout{** Please run "bibtex \jobname" to optain}
  \typeout{** the bibliography and then re-run LaTeX}
  \typeout{** twice to fix the references!}
  \typeout{******************************************}
  \typeout{}
 }

\end{document}